\documentclass[reprint,nofootinbib,amsmath,amssymb,aps,floatfix,superscriptaddress,twocolumn]{revtex4-2}
\usepackage{graphicx}
\usepackage{dcolumn}
\usepackage[colorlinks,linkcolor=magenta,anchorcolor=cyan,citecolor=blue]{hyperref}
\usepackage{bm}
\usepackage[multiple]{footmisc}

\begin{document}
\title{Shadow of Rotating Black Hole Immersed in Dark Matter Halo}

\author{Yong-You Liu}
\author{Tian-Yun Ye}

\author{Zhen Li} 
\email{Corresponding author: zhen.li@just.edu.cn}

\affiliation{School of Science, Jiangsu University of Science and Technology, Zhenjiang 212100, China}

\date{\today}

\begin{abstract}

The recent black hole shadow observations by the Event Horizon Telescope provide a unique opportunity to probe both the strong gravity regime and the environment surrounding supermassive black holes. Since dark matter halos are expected to exist around galactic-center black holes, they may leave observable signatures on black hole shadows. In this work, we investigate the spacetime structure, photon dynamics, and shadow properties of a rotating black hole immersed in a generalized double power law dark matter halo. Starting from a static dark matter black hole solution, we employ the Newman--Janis algorithm to construct its rotating counterpart and obtain a Kerr-like metric with a dark matter-modified mass function. We analyze the horizon structure, ergosphere, and unstable spherical photon orbits, and find that the dark matter halo generally enlarges the horizon, shifts the photon shell outward, and modifies the black hole shadow. The effects become more pronounced for rapidly rotating black holes. Using the shadow observation of Sgr A*, we further constrain the dark matter halo parameter space and identify correlations among the halo parameters, black hole spin, and observer inclination angle. Our results suggest that black hole shadow observations provide a promising probe of dark matter distributions near supermassive black holes.

\end{abstract}
\maketitle

\section{Introduction}

The direct imaging of black hole shadows by the Event Horizon Telescope (EHT) Collaboration has opened a new observational window into the strong gravity regime of general relativity. The first image of M87* \cite{shadow1} and the subsequent shadow observation of Sgr A* \cite{shadow2} have provided unprecedented opportunities to test gravitational theories in the vicinity of black holes and to probe the physical environment surrounding them. Unlike traditional electromagnetic observations that mainly reveal the properties of accretion flows, black hole shadows are directly determined by the spacetime geometry through the properties of unstable photon orbits \cite{shadow3}. Consequently, shadow observations have become a powerful tool for studying black hole physics, testing gravity theories, and exploring possible deviations from the standard Kerr paradigm.\cite{shaduse1,shaduse2,shaduse3,shaduse4,shaduse5}

At the same time, the nature of dark matter remains one of the most important unsolved problems in modern physics and cosmology. A wide range of astrophysical and cosmological observations, including galactic rotation curves \cite{rc}, gravitational lensing \cite{lens1,lens2}, galaxy cluster dynamics \cite{bullc1,bullc2}, and cosmic microwave background measurements \cite{cmb}, strongly support the existence of a non-luminous matter component that dominates the matter content of the Universe. However, despite extensive experimental and theoretical efforts, the fundamental properties of dark matter particles remain unknown. Since dark matter interacts predominantly through gravity, understanding its gravitational effects on astrophysical systems has become an important approach for investigating its nature.

Dark matter is generally believed to form extended halos around galaxies and galaxy clusters. Since supermassive black holes are located at the centers of galaxies, they are naturally embedded within dark matter environments. As a result, the surrounding dark matter halo may modify the spacetime geometry around the black hole and leave observable signatures in strong-gravity phenomena \cite{darkob}. In particular, black hole shadows provide a unique possibility to probe dark matter distributions in the immediate vicinity of galactic centers, complementing conventional observations that are typically sensitive only to much larger scales.

Theoretical studies of dark matter halos often rely on density profiles obtained from collisionless $N$-body simulations \cite{nby1,nby2}. Various phenomenological models have been proposed, including the Navarro--Frenk--White (NFW) profile \cite{nfw}, the Hernquist profile \cite{heq}, and many of their extensions \cite{prof1,prof2,prof3,prof4,prof5,prof7,prof8,prof9, zhao}. Among these models, the generalized Zhao profile \cite{zhao} provides a particularly useful framework because it contains several commonly used dark matter density distributions as special cases \cite{zhao1}. The Zhao profile is characterized by three shape parameters $(\alpha,\beta,\gamma)$, which respectively control the transition behavior, outer density slope, and inner density cusp of the halo. Owing to its flexibility, it offers a unified description of a broad class of dark matter distributions.

Motivated by this framework, in our previous work we constructed a static and spherically symmetric black hole solution immersed in a Zhao dark matter halo and investigated its observational properties \cite{zli}. We found that the dark matter halo can significantly modify the spacetime geometry, alter the photon sphere structure, and produce observable changes in the black hole shadow. Furthermore, by comparing the theoretical predictions with the EHT observations of Sgr A*, we obtained constraints on the parameter space of the dark matter halo model. These results demonstrated that black hole shadows can serve as an effective probe of dark matter distributions in galactic centers.

However, realistic astrophysical black holes are expected to possess nonzero angular momentum. Kerr solution is generally regarded as the standard description of astrophysical black holes in general relativity. Black hole rotation introduces a variety of new phenomena, including frame dragging, ergospheres, asymmetric photon regions, and distorted shadow shapes \cite{rosha1,rosha2}. Consequently, both the size and morphology of the shadow become strongly dependent on the spin parameter and the observer inclination angle. Since these effects are absent in static spacetimes, extending the dark matter black hole solution to the rotating case is essential for obtaining realistic observational predictions.

The primary objective of the present work is therefore to construct the rotating counterpart of the static dark matter black hole and investigate its physical and observational properties. To achieve this goal, we employ the Newman--Janis algorithm \cite{nj1,nj2}, which provides a well-established procedure for generating rotating solutions from static and spherically symmetric seed metrics. The resulting spacetime represents a rotating black hole embedded in a generalized Zhao dark matter halo and reduces smoothly to the Kerr solution in the absence of dark matter.

After obtaining the rotating metric, we systematically study its spacetime structure, including the horizons and ergosphere, and then analyze the motion of photons in this geometry. By determining the unstable spherical photon orbits, we derive the corresponding photon shell and calculate the black hole shadow for various dark matter halo configurations and spin parameters. We further investigate how the halo parameters modify the shadow size and explore the observational implications of these deviations. Finally, using the latest EHT observations of Sgr A* \cite{sgrdata}, we constrain the parameter space of the dark matter halo model and examine the degeneracies among the halo parameters, black hole spin, and observer inclination angle.

The remainder of this paper is organized as follows. In Sec.~\ref{sec2}, we construct the rotating dark matter black hole metric using the Newman--Janis method and analyze its spacetime structure, including the horizon and ergosphere properties. In Sec.~\ref{sec3}, we investigate the null geodesics, photon shell structure, and black hole shadow in this spacetime. In Sec.~\ref{sec4}, we employ the EHT shadow observations of Sgr A* to constrain the dark matter halo parameter space and discuss the corresponding observational implications. Finally, Sec.~\ref{sec5} summarizes our main results and presents future prospects.

\section{Rotating black hole in dark matter halo}\label{sec2}

\subsection{The Newman-Janis method}

We start from the static, spherically symmetric metric obtained from the Zhao density profile. The dark matter halo density is given by \cite{zhao}
\begin{equation}
\rho(r) = \rho_s \left( \frac{r}{r_s} \right)^{-\gamma} \left[ \left( \frac{r}{r_s} \right)^{\alpha} + 1 \right]^{(\gamma - \beta)/\alpha},
\end{equation}
where $\rho_s$ and $r_s$ denote the characteristic density and scale radius, respectively, while $\alpha$, $\beta$, and $\gamma$ control the shape of the profile. Specifically, $\gamma$ and $\beta$ respectively set the inner and outer slopes, and $\alpha$ determines the sharpness of the transition between these two regimes. All five parameters are nonnegative and can be adjusted to model a wide range of density distributions. This profile leads to the static metric \cite{zli}
\begin{equation}\label{static}
ds^2 = -f(r) dt^2 + \frac{dr^2}{f(r)} + h(r) \left( d\theta^2 + \sin^2\theta \, d\phi^2 \right),
\end{equation}
with $h(r)=r^2$ and
\begin{align}
f(r) &= 1 - \frac{2m(r)}{r},\\
m(r) &= M + \frac{4\pi\rho_s r_s^3}{\alpha}\, B\!\left( \frac{(r/r_s)^\alpha}{1 + (r/r_s)^\alpha};\, \frac{3-\gamma}{\alpha},\, \frac{\beta-3}{\alpha} \right).
\end{align}
Here $B(z; p,q)$ is the incomplete beta function, and $M$ is the central black hole mass.

To obtain the rotating counterpart, we apply the Newman-Janis algorithm \cite{nj1,nj2}. First, we transform the coordinates $(t,r,\theta,\phi)$ in Eq.~(\ref{static}) to Eddington-Finkelstein coordinates $(u,r,\theta,\phi)$ via
\begin{equation}
du = dt - \frac{dr}{f(r)}.
\end{equation}
In these coordinates the inverse metric can be expressed in terms of a null tetrad $(l^\mu, n^\mu, m^\mu, \bar m^\mu)$ as
\begin{equation}\label{tet}
g^{\mu\nu} = -l^\mu n^\nu - l^\nu n^\mu + m^\mu \bar m^\nu + m^\nu \bar m^\mu,
\end{equation}
with the tetrad vectors satisfying
\begin{align}
l^\mu l_\mu & = n^\mu n_\mu = m^\mu m_\mu = \bar m^\mu \bar m_\mu =0, \\
l^\mu m_\mu &= l^\mu \bar m_\mu = n^\mu m_\mu = n^\mu \bar m_\mu =0,\\
l^\mu n_\mu &= -m^\mu \bar m_\mu = -1.
\end{align}
A convenient choice is \cite{nj3,nj4}
\begin{align}
l^\mu &= \delta_r^\mu,\\
n^\mu &= \delta_u^\mu - \frac{f(r)}{2} \delta_r^\mu,\\
m^\mu &= \frac{1}{\sqrt{2h(r)}} \left( \delta_\theta^\mu + \frac{i}{\sin\theta} \delta_\phi^\mu \right),\\
\bar m^\mu &= \frac{1}{\sqrt{2h(r)}} \left( \delta_\theta^\mu - \frac{i}{\sin\theta} \delta_\phi^\mu \right).
\end{align}

The Newman-Janis method proceeds by performing a complex transformation in the $r$–$u$ plane:
\begin{equation}
r \to r' = r + i a \cos\theta, \qquad u \to u' = u - i a \cos\theta,
\end{equation}
where $a$ is the black hole spin, and leave $\theta$ and $\phi$ unchanged. After these transformations, the metric function also takes a new form 
\begin{equation}
f(r)\to \tilde f(r'),\quad h(r)\to \tilde h(r').
\end{equation}
In order to recover the Kerr solution in the limit of vanishing dark matter, we adopt the following transformation form of metric functions \cite{nj3,nj4}
\begin{align}
\tilde h(r') &= \Sigma \equiv r^2 + a^2\cos^2\theta,\label{ht}\\
\tilde f(r') &= 1 - \frac{2m(r) r}{\Sigma},\label{ft}
\end{align}
where the mass function $m(r)$ is assumed to remain a function of the real radial coordinate $r$ only. The tetrads are then transformed to \cite{nj3,nj4}
\begin{align}
l^{\prime\mu} &= \delta_r^\mu,\\
n^{\prime\mu} &= \delta_u^\mu - \frac{\tilde f(r')}{2} \delta_r^\mu,\\
m^{\prime\mu} &= \frac{1}{\sqrt{2\tilde h(r')}} \left( i a \sin\theta\,(\delta_u^\mu - \delta_r^\mu) + \delta_\theta^\mu + \frac{i}{\sin\theta} \delta_\phi^\mu \right),\\
\bar m^{\prime\mu} &= \frac{1}{\sqrt{2\tilde h(r')}} \left( -i a \sin\theta\,(\delta_u^\mu - \delta_r^\mu) + \delta_\theta^\mu - \frac{i}{\sin\theta} \delta_\phi^\mu \right).
\end{align}
Using this new tetrad, based on Eq.~(\ref{tet}), we obtain the metric in null coordinates:
\begin{align}
g_{uu} &= - \tilde f(r'), \quad g_{ur}=g_{ru}=-1,\\
g_{u\phi} &= g_{\phi u} = a\sin^2\theta\bigl(\tilde f(r')-1\bigr),\\
g_{r\phi} &= g_{\phi r} =  a\sin^2\theta,\\
g_{\theta\theta} &= \tilde h(r'),\\
g_{\phi\phi} &=\sin^2\theta\Bigl[\tilde h(r') + a^2\sin^2\theta\bigl(2-\tilde f(r')\bigr)\Bigr].
\end{align}

To express the metric in the familiar Boyer–Lindquist form, we perform the coordinate transformation
\begin{equation}
du = dt + F(r)\,dr,\qquad d\phi  = d\varphi + G(r)\,dr,
\end{equation}
where $F(r)$ and $G(r)$ can be determined by imposing the condition that all off-diagonal components of the metric tensor vanish except for the $t\varphi$ and $\varphi t$ components. Following Azreg-A{\"\i}nou prescription, they are given by \cite{nj3,nj4}
\begin{align}
F(r) &= - \frac{h(r) + a^2}
{f(r)h(r) + a^2},\\
G(r) &= - \frac{a}
{f(r)h(r) + a^2}.
\end{align}
After coordinate transformation, the non-vanishing metric components become
\begin{align}
g_{tt} &= - \tilde f(r'),\\
g_{t\varphi} &= g_{\varphi t} = a\sin^2\theta\bigl(\tilde f(r')-1\bigr),\\
g_{rr} &= \frac{\tilde h(r')}
{h(r)f(r) + a^2},\\
g_{\theta\theta} &= \tilde h(r'),\\
g_{\varphi \varphi} &= \sin^2\theta\Bigl[\tilde h(r') + a^2\sin^2\theta\bigl(2-\tilde f(r')\bigr)\Bigr].
\end{align}

% 新版
% g_{tt} &= - \tilde f(r'),\\
% g_{t\varphi} &= g_{\varphi t} = ah(r)\sin^2\theta \bigl( \frac{f(r)-1}{\tilde h(r')}\bigr),\\
% g_{rr} &= \frac{\tilde h(r')}
% {h(r)f(r) + a^2},\\
% g_{\theta\theta} &= \tilde h(r'),\\
% g_{\varphi \varphi} &= \sin^2\theta\Bigl[\tilde h(r') + a^2\sin^2\theta \frac{(2-f(r))h(r)+a^2\cos^2\theta}{\tilde h(r')}\Bigr].

After substituting Eq.(\ref{ht}) and (\ref{ft}), the line element in Boyer–Lindquist coordinates can be written as
\begin{equation}\label{metric}
ds^2 = g_{tt}\,dt^2 + g_{rr}\,dr^2 + g_{\theta\theta}\,d\theta^2 + g_{\varphi\varphi}\,d\varphi^2 + 2g_{t\varphi}\,dt\,d\varphi,
\end{equation}
where 
\begin{align}
g_{tt} &= -\left[\frac{\Delta - a^2\sin^2\theta}{\Sigma}\right],\\
g_{rr} &= \frac{\Sigma}{\Delta},\\
g_{t\varphi} &= -a\sin^2\theta\left[1 - \frac{\Delta - a^2\sin^2\theta}{\Sigma}\right],\\
g_{\theta\theta} &= \Sigma,\\
g_{\varphi \varphi} &= \sin^2\theta\left[\Sigma + a^2\sin^2\theta\left(2 - \frac{\Delta - a^2\sin^2\theta}{\Sigma}\right)\right],
\end{align}
and we have introduced
\begin{equation}
\Delta = r^2 + a^2 - 2m(r)r.
\end{equation}
The only deviation from the Kerr metric resides in the function $\Delta$, which now contains the modified mass function $m(r)$ arising from the dark matter halo. 

The metric (\ref{metric}) obtained through the Newman–Janis procedure should be regarded as an effective rotating extension of the static dark matter black hole. The corresponding stress-energy tensor generally describes a rotating anisotropic matter distribution and reduces to the original static dark matter source in the limit $a\to0$. A detailed reconstruction of the matter source is beyond the scope of the present work.

\subsection{Spacetime structure of the rotating dark matter black hole}

\begin{figure}
    \centering
    \includegraphics[scale=0.55]{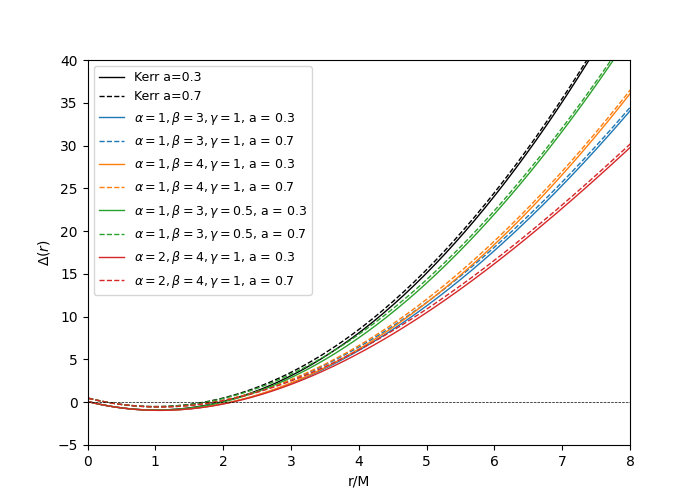}
    \caption{$\Delta(r)$ as a function of radius $r/M$, for different combinations of halo parameters $\alpha,\beta,\gamma$=(1,3,1), (1,4,1), (1,3,0.5), (2,4,1) and black hole spins $a/M=0.3, 0.7$, while we set $\rho_sM^2=0.0001$ and $r_s/M=30$. The Kerr case is also shown as a reference.}
\label{delta}
\end{figure}

\begin{figure}
    \centering
    \includegraphics[scale=0.55]{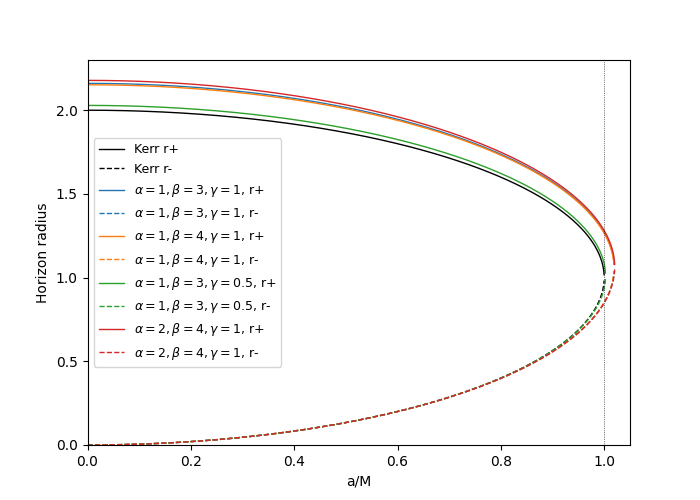}
    \caption{The horizons $r_\pm$ as a function of black hole spin $a/M$, with different combinations of halo parameters $\alpha,\beta,\gamma$=(1,3,1), (1,4,1), (1,3,0.5), (2,4,1), while we set $\rho_sM^2=0.0001$ and $r_s/M=30$. The Kerr case is also shown as a reference. The solid and dashed curves represent the outer event horizon and inner horizon, respectively.}
\label{horz}
\end{figure}

\begin{figure*}
    \centering
    \includegraphics[scale=0.5]{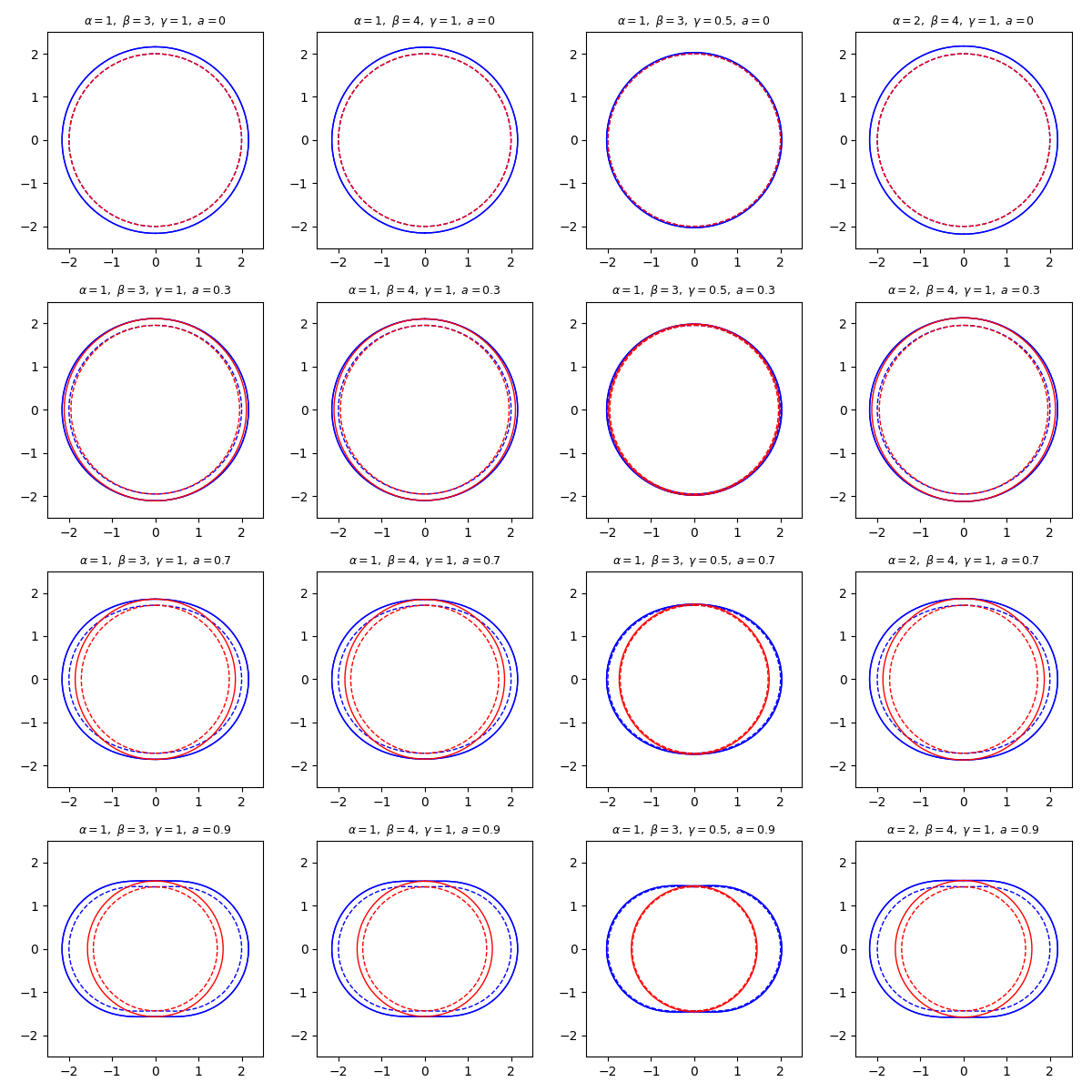}
    \caption{The shape of the ergosphere and event horizon for rotating dark matter black hole. The coordinates in each subplot are given in units of $M$. The solid blue and dashed blue curves represent the ergosphere of rotating dark matter black hole and Kerr black hole, respectively. The solid red and dashed red curves are the horizon of rotating dark matter black hole and Kerr black hole, respectively. We present the results for several combinations of the halo parameters $\alpha,\beta,\gamma$=(1,3,1), (1,4,1), (1,3,0.5), (2,4,1), together with black hole spins $a/M=0, 0.3, 0.7, 0.9$. Throughout this analysis, the halo parameters are fixed at $\rho_sM^2=0.0001$ and $r_s/M=30$. }
\label{ergo}
\end{figure*}

The global properties of the rotating dark matter black hole are completely encoded in the metric function
\begin{equation}
\Delta(r)=r^{2}+a^{2}-2m(r)r,
\end{equation}
where the dark matter halo modifies the effective mass function $m(r)$. In the Kerr spacetime, the mass parameter is a constant, $m(r)=M$, whereas in the present case the accumulated dark matter mass enclosed within radius $r$ contributes to the gravitational field and introduces a nontrivial radial dependence. Consequently, all horizon-related properties of the spacetime are modified through the function $\Delta(r)$.

The locations of the horizons are determined by the roots of
\begin{equation}
\Delta(r)=0.
\end{equation}
Depending on the values of the spin parameter and the dark matter halo parameters, the equation may admit two distinct real roots, a degenerate root, or no real roots. These cases correspond respectively to a non-extremal black hole, an extremal black hole, and a naked singularity. Therefore, the behavior of $\Delta(r)$ provides direct information about the causal structure of the spacetime.

Fig.~\ref{delta} displays $\Delta(r)$ as a function of the radial coordinate for several representative dark matter halo profiles characterized by $(\alpha,\beta,\gamma)=(1,3,1)$, $(1,4,1)$, $(1,3,0.5)$, and $(2,4,1)$, together with two spin values $a/M=0.3$ and $0.7$. For comparison, the corresponding Kerr solution is also shown. Throughout this section we fix the halo density and scale radius to $\rho_s M^2=10^{-4}$ and $r_s/M=30$.

Several important features can be observed from Fig.~\ref{delta}. In the vicinity of the black hole, the curves associated with different halo profiles almost coincide with the Kerr solution. This behavior is expected because the dark matter contribution enclosed within the innermost region is relatively small compared with the black hole mass. As a result, the spacetime remains approximately Kerr-like near the horizon. However, as the radial coordinate increases, the cumulative contribution from the dark matter halo becomes increasingly important, leading to noticeable deviations from the Kerr behavior. In particular, the slope of $\Delta(r)$ at large radii is modified by the dark matter distribution, indicating that the asymptotic gravitational field differs from that of an isolated rotating black hole.

To investigate the horizon structure more systematically, we solve the equation $\Delta(r)=0$ numerically and determine the inner horizon $r_-$ and outer horizon $r_+$ as functions of the spin parameter. The results are shown in Fig.~\ref{horz}. The solid curves represent the outer event horizon, while the dashed curves correspond to the inner Cauchy horizon.

Several interesting trends emerge. First, the outer horizon is significantly more sensitive to the dark matter halo than the inner horizon. Physically, the outer horizon is located farther away from the black hole center and therefore encloses a larger amount of dark matter mass. Since the halo contribution accumulates with radius, its effect naturally becomes more pronounced at the outer horizon. By contrast, the inner horizon resides deep inside the strong-field region where the enclosed dark matter mass is comparatively small, resulting in only minor deviations from the Kerr prediction.

Second, the presence of dark matter shifts the extremal limit of the rotating black hole. In the Kerr spacetime, extremality occurs at $a/M=1$, where the two horizons merge into a single degenerate horizon. In the present model, however, the additional gravitational mass contributed by the dark matter halo modifies this condition. As shown in Fig.~\ref{horz}, the horizon structure can remain even for spin parameters slightly exceeding $a/M=1$. This does not imply a violation of cosmic censorship; rather, it reflects the fact that the total gravitating mass of the system is larger than the bare black hole mass $M$ due to the surrounding dark matter distribution. Consequently, the effective extremal spin parameter can exceed unity when expressed in terms of the central black hole mass alone.

Another characteristic feature of rotating black holes is the existence of an ergoregion. The outer boundary of this region, known as the stationary limit surface, is determined by the condition
\begin{equation}
g_{tt}=0.
\end{equation}
The ergoregion plays a fundamental role in energy extraction processes such as the Penrose mechanism and superradiant scattering.

The shapes of the horizon and ergosphere are illustrated in Fig.~\ref{ergo} for several halo profiles and spin values $a/M=0$, $0.3$, $0.7$, and $0.9$. The corresponding Kerr case are displayed for comparison. For the non-rotating case, the horizon and stationary limit surface coincide, and no ergoregion exists. As the spin increases, the stationary limit surface gradually separates from the horizon, producing the familiar ergosphere structure.

The influence of the dark matter halo becomes increasingly apparent at higher spins. Compared with the Kerr spacetime, the ergosphere of the rotating dark matter black hole is slightly enlarged and deformed. This modification originates from the dark matter contribution to $\Delta(r)$, which changes both the horizon radius and the location of the stationary limit surface. Similar to the horizon analysis, the strongest deviations are generated by variations of the inner slope parameter $\gamma$. Halo profiles with larger values of $\gamma$ possess a steeper central density distribution and therefore induce a stronger modification of the near-horizon geometry. On the other hand, changes in $\alpha$ and $\beta$ mainly affect the outer halo structure and lead to comparatively weaker effects.

The enlargement of the ergoregion has potentially important astrophysical implications. Since the efficiency of rotational energy extraction depends sensitively on the size and geometry of the ergosphere, the presence of a dark matter halo may influence jet formation, accretion dynamics, and superradiant phenomena around rapidly rotating black holes. Although a detailed investigation of these processes lies beyond the scope of the present work, the modified ergosphere geometry suggests that observable signatures may arise in realistic astrophysical environments.

In summary, the Newman--Janis construction yields a rotating dark matter black hole spacetime that preserves the essential features of the Kerr geometry while incorporating the gravitational effects of a surrounding dark matter halo through the modified mass function $m(r)$. The resulting deviations alter the horizon locations, shift the extremal spin bound, and modify the shape of the ergosphere. These changes directly affect the null geodesic structure of the spacetime and are expected to leave observable imprints on black hole shadows and other strong-gravity phenomena, which will be investigated in the following sections.

\section{Null geodesics and shadow}\label{sec3}

\begin{figure}
    \centering
    \includegraphics[scale=0.55]{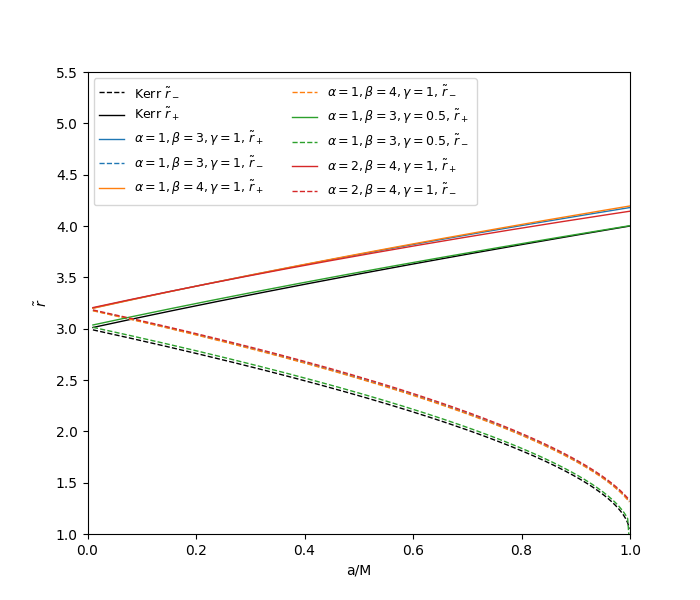}
    \caption{The inner (dashed curves, ${\tilde r}_-$) and outer (solid curves, ${\tilde r}_+$) boundary of photon shell as a function of black hole spin $a/M$ for rotating dark matter black hole, with different combinations of halo parameters $\alpha,\beta,\gamma$=(1,3,1), (1,4,1), (1,3,0.5), (2,4,1), while we set $\rho_sM^2=0.0001$ and $r_s/M=30$. The Kerr case is also shown as a reference. }
\label{phs}
\end{figure}

We now study the motion of massless particles (photons) in the rotating dark matter black hole spacetime. The metric (\ref{metric}) retains the Kerr-like structure and admits separability of the Hamilton–Jacobi equation, similar to many rotating black hole solutions generated via the Newman–Janis procedure. This spacetime is still stationary and axisymmetric, admitting two Killing vectors $\partial_t$ and $\partial_\phi$, leading to the conserved specific energy $E$ and angular momentum $L$ about the symmetry axis. In addition, the separability of the Hamilton–Jacobi equation for null geodesics gives a third constant of motion, the Carter constant $Q$ \cite{kerrsym}. It is convenient to introduce the energy-scaled quantities
\begin{equation}
\lambda = L/E, \qquad \eta = Q/E^2.
\end{equation}

In Boyer–Lindquist coordinates, the null geodesic equations take the form
\begin{align}
\frac{\Sigma}{E} p^{r} &= \pm_{r} \sqrt{\mathcal{R}(r)}, \\
\frac{\Sigma}{E} p^{\theta} &= \pm_{\theta} \sqrt{\Theta(\theta)}, \\
\frac{\Sigma}{E} p^{\varphi} &= \frac{a}{\Delta}\left(r^{2}+a^{2}-a\lambda\right) + \frac{\lambda}{\sin^{2}\theta} - a, \\
\frac{\Sigma}{E} p^{t} &= \frac{r^{2}+a^{2}}{\Delta}\left(r^{2}+a^{2}-a\lambda\right) + a\left(\lambda - a\sin^{2}\theta\right),
\end{align}
where $\Sigma = r^{2}+a^{2}\cos^{2}\theta$, $\Delta = r^{2}+a^{2}-2m(r)r$, and the radial and angular potentials are defined as
\begin{align}
\mathcal{R}(r) &= \left(r^{2}+a^{2}-a\lambda\right)^{2} - \Delta\left[\eta + (\lambda - a)^{2}\right], \\
\Theta(\theta) &= \eta + a^{2}\cos^{2}\theta - \lambda^{2}\cot^{2}\theta.
\end{align}

Circular null geodesics (photon spheres) correspond to unstable orbits that satisfy both $\mathcal{R}(r)=0$ and $\mathcal{R}'(r)=0$ at some radius $\tilde r$ outside the outer horizon $r_+$. These conditions determine the critical values of the conserved quantities $\lambda$ and $\eta$ as functions of $\tilde r$:
\begin{align}
\tilde\lambda &= a + \frac{\tilde r}{a}\left[\tilde r - \frac{2\tilde\Delta}{\tilde r - M - N\tilde r}\right]\label{lam}, \\
\tilde\eta &= \frac{\tilde r^{3}}{a^{2}}\left[\frac{4\tilde\Delta\,(M - N\tilde r)}{(\tilde r - M - N\tilde r)^{2}} - \tilde r\right]\label{eta},
\end{align}
where
\begin{equation}
\tilde\Delta = \Delta(\tilde r) = \tilde r^{2}+a^{2}-2m(\tilde r)\tilde r,\qquad
N = \left.\frac{dm(r)}{dr}\right|_{r=\tilde r}.
\end{equation}
The derivative $N$ encodes the dark matter halo density at the critical radius. For a given spin $a$ and dark matter parameters, the condition $\tilde\eta = 0$ yields two radii $\tilde r_-$ and $\tilde r_+$ ($\tilde r_- < \tilde r_+$) which define the inner and outer boundaries of the photon shell – the region in which spherical photon orbits exist. Fig.~\ref{phs} illustrates the inner ($\tilde r_-$) and outer ($\tilde r_+$) boundaries of the photon shell as functions of the black hole spin parameter for several representative dark matter halo profiles,
$(\alpha,\beta,\gamma)=(1,3,1)$, $(1,4,1)$, $(1,3,0.5)$, and $(2,4,1)$, while the halo density and scale radius are fixed to
$\rho_s M^2=10^{-4}$ and $r_s/M=30$, respectively.
For comparison, the corresponding Kerr results are also shown.

The photon shell represents the set of unstable spherical photon orbits surrounding the black hole and plays a central role in determining the observed shadow. Photons with impact parameters close to the critical values $(\tilde\lambda,\tilde\eta)$ can orbit the black hole multiple times before eventually escaping to a distant observer or plunging into the horizon. Consequently, the location and extent of the photon shell encode the fundamental properties of the underlying spacetime geometry and directly determine the size and morphology of the black hole shadow.

From Fig.~\ref{phs}, we can observe that both the inner and outer boundaries of the photon shell are shifted outward relative to the Kerr case. This behavior reflects the additional gravitational field generated by the dark matter halo. Since the effective mass function $m(r)$ increases with radius, the spacetime curvature experienced by photons becomes stronger than that of an isolated Kerr black hole. As a result, unstable spherical photon orbits can exist at larger radii, leading to an expansion of the entire photon shell.

Another important observation is that the parameter $\gamma$ produces the most substantial modification of the photon shell. The reason can be traced to its role in controlling the inner density cusp of the dark matter halo. Since the photon shell is located only a few gravitational radii outside the horizon, its properties are mainly determined by the dark matter distribution in the innermost region of the halo. A larger value of $\gamma$ corresponds to a steeper central density profile and therefore a stronger gravitational contribution near the photon shell. In contrast, the parameters $\alpha$ and $\beta$ mainly govern the transition and outer regions of the halo profile and thus generate comparatively weaker effects on the photon shell structure.

The outward displacement of the photon shell has direct observational consequences. Since the shadow boundary is generated by photons asymptotically approaching these unstable spherical orbits, a larger photon shell generally implies a larger shadow size. Therefore, the modifications shown in Fig.~\ref{phs} already suggest that the presence of dark matter can produce measurable deviations in black hole shadow observations.

The observable shadow is determined by projecting the critical photon trajectories onto the celestial plane of a distant observer. For a distant observer located at inclination angle $\theta_0$, the apparent position of a photon on the celestial plane is described by the coordinates $X$ (horizontal displacement) and $Y$ (vertical displacement). These are related to the conserved quantities $\lambda$ and $\eta$ via the Bardeen relations \cite{bar}:
\begin{equation}
X = -\frac{\lambda}{\sin\theta_0}, \qquad 
Y = \pm \sqrt{\eta + a^2\cos^2\theta_0 - \lambda^2\cot^2\theta_0}.
\end{equation}
where $\pm$ is the sign of $\cos\theta_0$. By substituting the critical impact parameters $\tilde\lambda(\tilde r)$ and $\tilde\eta(\tilde r)$ of Eq.(\ref{lam}) and (\ref{eta}) into these relations and allowing $\tilde r$ to vary continuously across the photon shell interval $[\tilde r_-,\tilde r_+]$, one obtains the complete critical curve, namely the shadow boundary.

\begin{figure*}
    \centering
    \includegraphics[scale=0.6]{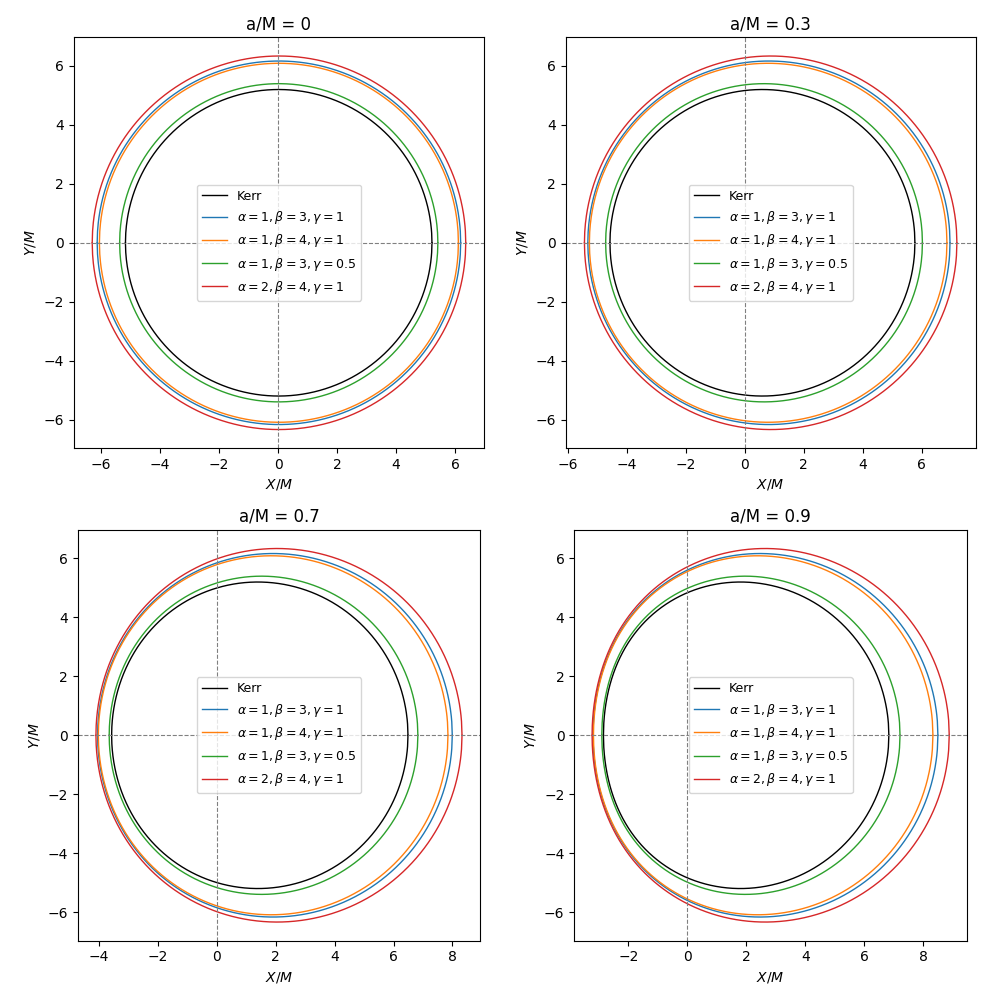}
    \caption{The shadow of the rotating dark matter black hole as viewed by an equatorial observer at spatial infinity, for different combinations of halo parameters $\alpha,\beta,\gamma$=(1,3,1), (1,4,1), (1,3,0.5), (2,4,1) and black hole spins $a/M=0, 0.3, 0.7, 0.9$, while we set $\rho_sM^2=0.0001$ and $r_s/M=30$. The Kerr case is also shown as a reference. }
\label{shadow}
\end{figure*}

Fig.\ref{shadow} presents the resulting shadow shapes for different halo profiles and black hole spin parameters.
The corresponding Kerr shadows are displayed as references. Several characteristic features emerge from the figure.

For the non-rotating case ($a=0$), the shadow remains circular due to spherical symmetry. Nevertheless, the dark matter halo slightly enlarges the shadow radius relative to the Schwarzschild limit. This enlargement originates from the outward shift of the photon sphere induced by the additional gravitational potential of the halo.

As the spin increases, the shadow becomes progressively distorted due to frame dragging. The left-hand side of the shadow is compressed while the right-hand side is stretched, producing the characteristic D-shaped familiar from Kerr black holes. In the presence of dark matter, however, the shadow is not merely a rescaled Kerr shadow. Instead, both its overall size and detailed shape are modified because the dark matter halo changes the entire structure of the photon shell from which the shadow is formed.

The deviations become particularly visible for rapidly rotating black holes with $a/M=0.7$ and $0.9$. In these cases, the combined effects of frame dragging and the modified mass distribution lead to noticeable differences between the dark matter shadow and the Kerr prediction. The shadow boundary is shifted outward, and the asymmetry introduced by rotation is altered. Such deviations may become observable with future high-resolution very-long-baseline interferometry observations.

Consistent with the results obtained for the horizon, ergosphere, and photon shell, the parameter $\gamma$ again produces the largest modification of the shadow morphology. Since $\gamma$ controls the dark matter density near the black hole, it directly affects the unstable photon orbits that generate the shadow boundary. By contrast, variations in $\alpha$ and $\beta$ mainly influence the outer halo structure and therefore have a comparatively weaker impact on the shadow. Nevertheless, their effects are not entirely negligible. Unlike horizon-related quantities, shadow-forming photons can travel through an extended region of the spacetime before reaching the observer. Consequently, the entire dark matter halo contributes to the photon trajectories to some extent, allowing even the outer halo parameters to leave small but potentially detectable imprints on the shadow shape. This establishes black hole shadow observations as a powerful tool for probing the dark matter distribution in the vicinity of supermassive black holes and the dark matter halo of host galaxy.

\section{Observational constraints from Sgr A* shadow data}\label{sec4}

\begin{figure*}
    \centering
    \includegraphics[scale=0.4]{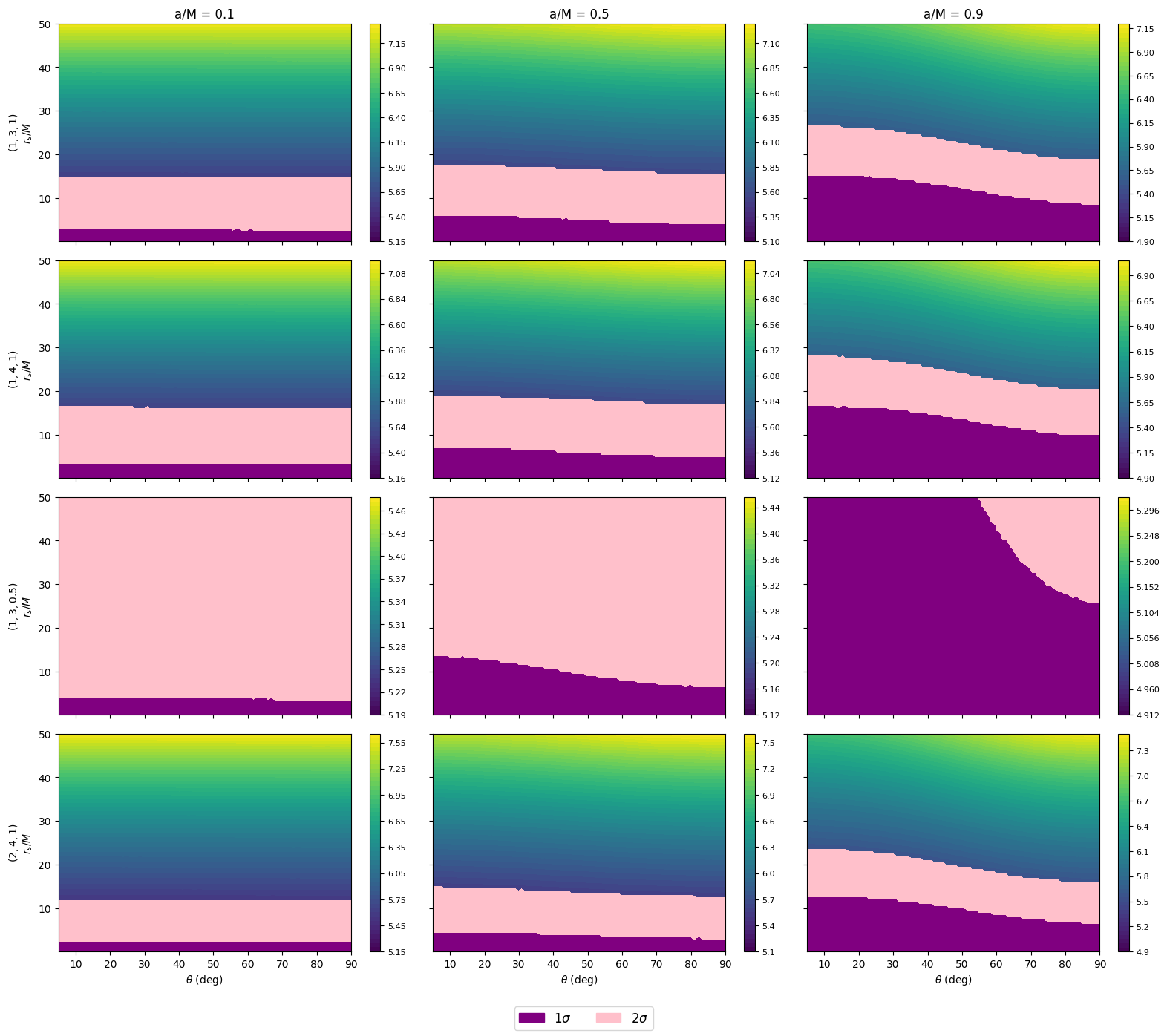}
    \caption{Constraints from the Sgr A* shadow observations in the parameter space of $r_s/M$ and observational angle $\theta$ for different halo parameter combinations $(\alpha, \beta, \gamma) = (1, 3, 1), (1, 4, 1), (1, 3, 0.5), (2, 4, 1)$ and black hole spins $a/M=0.1, 0.5, 0.9$, while we fix the $\rho_s= 0.0001M^2$. The color bar on the right of each subplot indicates the average shadow radius $\bar R_s$. The pink and purple regions correspond to the $2\sigma$ and $1\sigma$ confidence regions of the Sgr A* shadow observations, respectively.}
\label{fixrho}
\end{figure*}

\begin{figure*}
    \centering
    \includegraphics[scale=0.4]{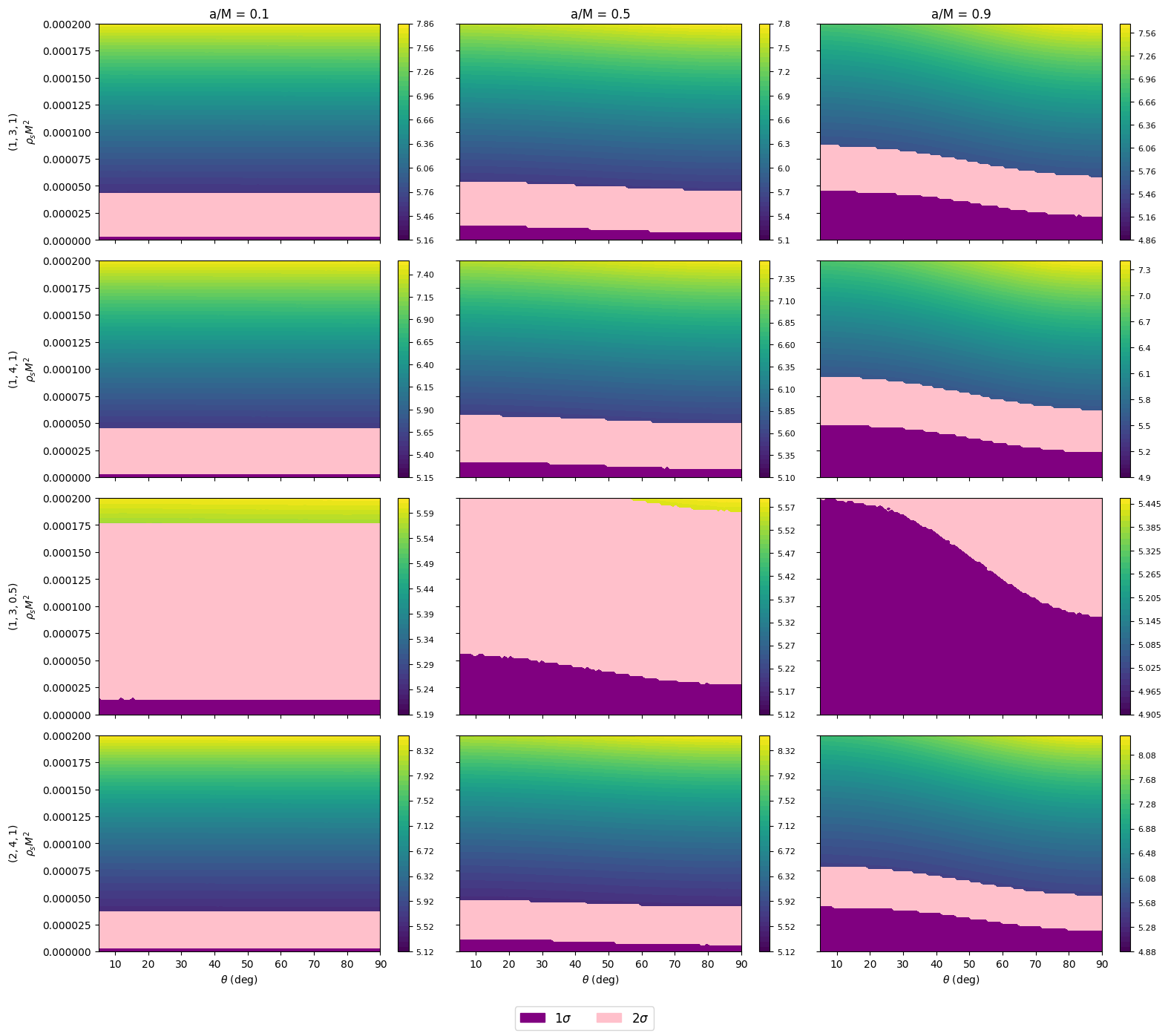}
    \caption{Constraints from the Sgr A* shadow observations in the parameter space of $\rho_s M^2$ and observational angle $\theta$ for different halo parameter combinations $(\alpha, \beta, \gamma) = (1, 3, 1), (1, 4, 1), (1, 3, 0.5), (2, 4, 1)$ and black hole spins $a/M=0.1, 0.5, 0.9$, while we fix the $r_s=30M$. The color bar on the right of each subplot indicates the average shadow radius $\bar R_s$. The pink and purple regions correspond to the $2\sigma$ and $1\sigma$ confidence regions of the Sgr A* shadow observations, respectively.}
\label{fixrs}
\end{figure*} 

\begin{figure*}
    \centering
    \includegraphics[scale=0.4]{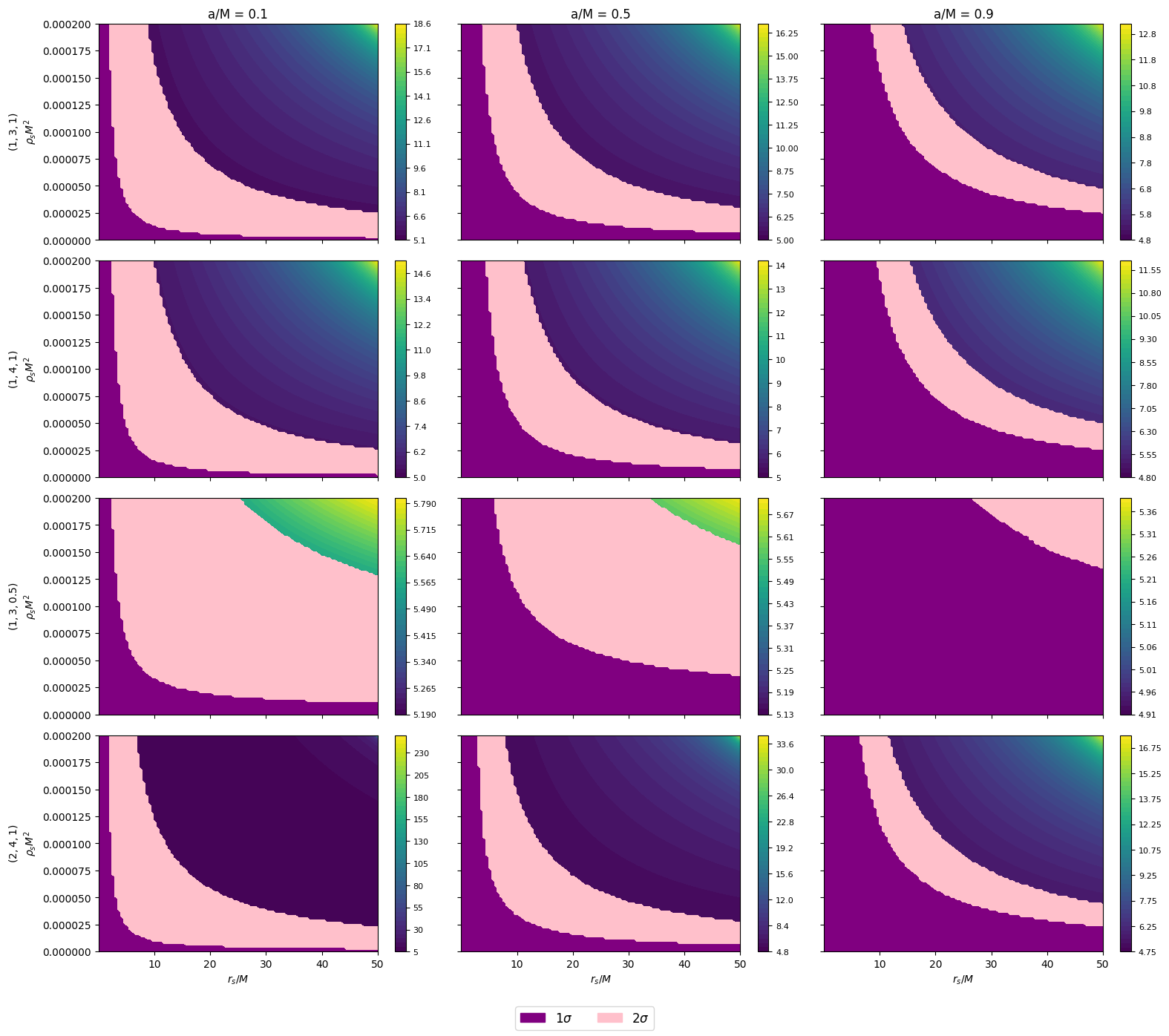}
    \caption{Constraints from the Sgr A* shadow observations in the parameter space of $\rho_s M^2$ and $r_s/M$ for different halo parameter combinations $(\alpha, \beta, \gamma) = (1, 3, 1), (1, 4, 1), (1, 3, 0.5), (2, 4, 1)$ and black hole spins $a/M=0.1, 0.5, 0.9$, while we fix the observational angle $\theta=30^\circ$. The color bar on the right of each subplot indicates the average shadow radius $\bar R_s$. The pink and purple regions correspond to the $2\sigma$ and $1\sigma$ confidence regions of the Sgr A* shadow observations, respectively.}
\label{fixtheta}
\end{figure*}

The black hole shadow provides a unique probe of the spacetime geometry in the strong-field regime and therefore offers an opportunity to test the influence of dark matter distributions surrounding supermassive black holes. Since the shadow size is primarily determined by the structure of the photon region, any modification of the spacetime metric induced by a dark matter halo will leave observable imprints on the shadow morphology. Consequently, the recent shadow observations of Sgr A* by EHT collaboration can be used to constrain the parameters of the dark matter halo model considered in this work.

Among currently available black hole shadow observations, Sgr A* is particularly interesting because it resides at the center of the Milky Way, whose dark matter distribution has been extensively studied through both astrophysical observations and direct detection experiments. Therefore, constraints obtained from the shadow of Sgr A* can be directly connected to the dark matter environment in our Galaxy. Furthermore, compared with M87* \cite{m87data}, the shadow size of Sgr A* is measured with relatively high precision \cite{sgrdata}, making it a suitable target for testing deviations from the Kerr geometry induced by dark matter halos.

The observed angular diameter of the Sgr A* shadow can be translated into shadow radius. Following the EHT analysis, the allowed $1\sigma$ and $2\sigma$ ranges of the shadow radius are taken to be \cite{sgrdata}
\begin{equation}
R_{\rm obs}^{1\sigma} \in [4.55,\,5.22]M,
\end{equation}
and
\begin{equation}
R_{\rm obs}^{2\sigma} \in [4.21,\,5.56]M.
\end{equation}
For rotating black holes, the shadow is generally non-circular and exhibits a noticeable asymmetry due to frame dragging. As a result, the shadow radius depends on the polar angle measured on the observer's celestial plane. To facilitate comparison with observational data, we define the average shadow radius as
\begin{equation}
\bar R_s=\frac{1}{N}\sum_{i=1}^{N}\sqrt{X_i^2+Y_i^2},
\end{equation}
where $(X_i, Y_i)$ denote the coordinates of the i-th sampled point on the critical curve, and N is the total number of sampling points (we set N=800 here, We have verified that the value of $\bar R_s$ converges as N increases, and further refinement of the sampling resolution produces negligible changes in the results). This averaged shadow radius characterizes the overall size of the shadow while remaining insensitive to local distortions of the boundary.

Another important factor is the observer inclination angle $\theta_o$. Since the apparent shadow shape varies significantly with viewing geometry, especially for rapidly rotating black holes, the inclination angle must also be treated as a free parameter. In the following analysis, we therefore constrain the dark matter halo parameters jointly with the observer inclination angle by requiring the predicted averaged shadow radius to lie within the observationally allowed ranges.

Fig.~\ref{fixrho} presents the observational constraints in the two-dimensional parameter space $(r_s,\theta_o)$ while fixing the characteristic halo density to $\rho_s M^2 = 10^{-4}$. Different rows correspond to the halo profiles $(\alpha,\beta,\gamma)=(1,3,1)$, $(1,4,1)$, $(1,3,0.5)$, and $(2,4,1)$, whereas the columns represent black hole spins $a/M=0.1$, $0.5$, and $0.9$, respectively. The color map in each panel indicates the predicted average shadow radius, while the purple and pink regions denote the observationally allowed $1\sigma$ and $2\sigma$ intervals.

Several important features can be identified. First, the observation allowed parameter region becomes significantly larger as the black hole spin increases. Second, the dependence on the inclination angle becomes progressively stronger for larger spins. For slowly rotating black holes, the shadow remains nearly circular and is only weakly affected by the observer orientation. In contrast, for near-extremal spins, the shadow distortion is substantial and the viewing angle plays a crucial role in determining the average shadow size.

More importantly, the parameter $\gamma$, which controls the inner density slope of the dark matter halo, produces the most significant influence on the observational constraints. Comparing the panels with $\gamma=1$ and $\gamma=0.5$, one observes a systematic shift of the allowed regions toward smaller values of $r_s$ for steeper inner density profiles. This behaviour can be understood from the fact that a larger $\gamma$ concentrates more dark matter in the vicinity of the black hole, thereby producing stronger modifications to the photon shell and enlarging the shadow radius. To remain compatible with observations, the characteristic scale radius must decrease correspondingly. In contrast, variations of $\alpha$ and $\beta$, which mainly control the transition and outer slope of the halo profile, generate only minor changes in the allowed regions. This is expected because the shadow is predominantly determined by photon trajectories in the strong-gravity region close to the black hole, where the inner halo structure is most relevant.

Fig.~\ref{fixrs} shows the complementary constraints in the $(\rho_s,\theta_o)$ plane with the scale radius fixed at $r_s/M=30$. Similar trends emerge. The allowed parameter regions expand as the spin increases, and the sensitivity to the observer inclination becomes more pronounced for rapidly rotating black holes. In addition, a clear correlation between the characteristic density $\rho_s$ and the inner slope parameter $\gamma$ can be observed. Models with steeper inner density cusps generally require lower values of $\rho_s$ to satisfy the observational constraints. This degeneracy arises because both $\rho_s$ and $\gamma$ enhance the effective dark matter mass enclosed within the photon region. Increasing either parameter tends to enlarge the shadow size. Consequently, a larger $\gamma$ must be compensated by a smaller $\rho_s$ in order to reproduce the observed shadow radius.

The weak dependence on $\alpha$ and $\beta$ again confirms that the EHT shadow measurement primarily probes the inner dark matter distribution rather than the outer halo structure. This result is particularly interesting because most conventional astrophysical observations, such as galactic rotation curves, are sensitive to much larger spatial scales. Black hole shadows therefore provide a complementary method for investigating dark matter distributions in the central regions of galaxies.

In Fig.~\ref{fixtheta}, we fix the observer inclination angle at $\theta_o=30^\circ$ and explore the joint constraints in the $(\rho_s,r_s)$ parameter space. The resulting contours reveal a pronounced anti-correlation between the characteristic density and scale radius. Specifically, larger values of $\rho_s$ require smaller values of $r_s$, whereas smaller densities can be compensated by larger scale radii. This behaviour reflects the fact that the shadow is mainly sensitive to the total dark matter contribution near the photon shell. Different combinations of $\rho_s$ and $r_s$ can generate a similar enclosed mass in this region, leading to a degeneracy in the observational constraints.

The influence of the spin parameter is also clearly visible. For low-spin configurations, the allowed region is relatively narrow, indicating that the shadow size provides stronger constraints on the halo properties. As the spin approaches the extremal regime, the allowed parameter space broadens considerably because the rotational effects themselves become a major source of variation in the shadow radius. Consequently, uncertainties in the spin parameter can partially mask the impact of the dark matter halo.

Comparing different halo profiles further demonstrates that the inner slope parameter $\gamma$ remains the dominant quantity controlling the observational constraints. Larger values of $\gamma$ shift the allowed regions toward lower values of either $\rho_s$ or $r_s$, reflecting the stronger gravitational influence of cuspy dark matter distributions near the black hole. By contrast, the dependence on $\alpha$ and $\beta$ remains relatively weak throughout the parameter space explored here.

Overall, Fig.~\ref{fixrho}-\ref{fixtheta} consistently indicate that current Sgr A* shadow observations are capable of placing meaningful constraints on the dark matter distribution surrounding the Galactic center. Although substantial parameter degeneracies remain, particularly among $\rho_s$, $r_s$, and the black hole spin, the data already exhibit a clear preference for models with moderate dark matter contributions in the vicinity of the photon shell. Future improvements in shadow imaging accuracy, together with independent measurements of the black hole spin and inclination angle, are expected to significantly tighten these constraints and provide a powerful probe of the dark matter environment in the innermost region of the Milky Way.

\section{Conclusion}\label{sec5}

In this work, we have investigated the spacetime geometry, null geodesic structure, black hole shadow, and observational constraints of a rotating black hole immersed in a dark matter halo described by the generalized Zhao density profile. Starting from the corresponding static dark matter black hole solution, we employed the Newman--Janis algorithm to construct its rotating counterpart and obtained a Kerr-like spacetime in which all dark matter effects are encoded in the modified mass function $m(r)$. This framework allows us to systematically study how different dark matter distributions affect the near-horizon geometry and the observable properties of rotating black holes.

We first analyzed the spacetime structure of the rotating dark matter black hole. By studying the behavior of the metric function $\Delta(r)$, we found that the presence of the dark matter halo modifies the horizon structure relative to the Kerr spacetime. The deviations become increasingly significant at larger radii due to the cumulative contribution of the dark matter mass. Numerical calculations of the inner and outer horizons showed that the outer horizon is considerably more sensitive to the dark matter distribution than the inner horizon. We also found that the dark matter halo alters the extremality condition of the rotating black hole, allowing the critical spin parameter to exceed the Kerr bound $a/M=1$ when expressed in terms of the central black hole mass alone. Furthermore, the ergosphere is enlarged and deformed by the dark matter halo, implying potential modifications to energy extraction processes and other rotational phenomena occurring in the vicinity of the black hole.

Subsequently, we investigated the motion of photons in this spacetime by solving the null geodesic equations and deriving the corresponding photon shell structure. The unstable spherical photon orbits were obtained through the critical conditions imposed on the radial effective potential. Our results demonstrated that the entire photon shell is shifted outward compared with the Kerr case due to the additional gravitational field generated by the surrounding dark matter halo. The effect becomes increasingly pronounced for rapidly rotating black holes, indicating that high-spin systems provide a particularly sensitive probe of dark matter-induced modifications. Among the halo parameters, the inner density slope parameter $\gamma$ was found to have the strongest influence on the photon shell structure, whereas the parameters $\alpha$ and $\beta$ produce relatively smaller effects.

Based on the photon shell properties, we then constructed the apparent shadow of the rotating dark matter black hole as seen by a distant observer. We showed that the dark matter halo modifies both the size and shape of the shadow. For non-rotating black holes, the shadow remains circular but its radius increases due to the outward displacement of the photon sphere. For rotating black holes, the familiar Kerr-like asymmetry caused by frame dragging persists; however, the shadow boundary is noticeably shifted relative to the Kerr prediction. The deviations become more significant as the spin increases, reflecting the interplay between rotational effects and the dark matter induced modification of the spacetime geometry. Consistent with the horizon and photon shell analyses, the parameter $\gamma$ was identified as the dominant factor controlling the shadow deviations, while $\alpha$ and $\beta$ mainly generate subleading corrections.

To explore the observational implications of these results, we employed the shadow observations of Sgr A* to constrain the parameter space of the dark matter halo. Using the averaged shadow radius as the observational observable, we examined the observation allowed regions in the parameter spaces $(r_s,\theta_o)$, $(\rho_s,\theta_o)$, and $(\rho_s,r_s)$ for different halo profiles and black hole spins. Our analysis revealed that the observational constraints are strongly affected by both the black hole spin and the observer inclination angle. Higher spin configurations generally allow a broader parameter space due to the enhanced asymmetry of the shadow. We also found that the observational data are particularly sensitive to the inner density slope $\gamma$, whereas the dependence on $\alpha$ and $\beta$ remains relatively weak. In addition, a clear degeneracy between the characteristic halo density $\rho_s$ and scale radius $r_s$ was identified, indicating that different dark matter configurations may produce similar shadow sizes through comparable gravitational effects near the photon region.

Overall, our results demonstrate that black hole shadows provide a valuable and complementary probe of dark matter distributions in the immediate vicinity of supermassive black holes. Unlike traditional astrophysical observations, which primarily constrain dark matter on galactic or cosmological scales, shadow observations directly probe the strong gravity region where the interplay between dark matter and spacetime geometry becomes most significant. The modifications found in the horizon structure, ergosphere geometry, photon shell, and shadow morphology collectively suggest that future high precision black hole imaging observations may offer a novel method for investigating the nature and distribution of dark matter near galactic centers.

\section*{Acknowledgments}
This work is supported by the National Natural Science Foundation of China (Grant No.~12505073). Zhen Li acknowledges the financial support from the Start-up Funds for Doctoral Talents at Jiangsu University of Science and Technology. 
\\
\\

\end{document}